\documentclass{article}
\usepackage{spconf,amsmath,graphicx}
\usepackage{multirow}
\usepackage{caption}
\usepackage{hyperref}
\usepackage{flushend}

\title{Towards Sub-millisecond Latency Real-Time Speech Enhancement Models on Hearables}
\name{
    Artem Dementyev\textsuperscript{1},
    Chandan K. A. Reddy\textsuperscript{3},
    Scott Wisdom\textsuperscript{1},
    Navin Chatlani\textsuperscript{3}, 
}
\namesecondline{
    John R.\ Hershey\textsuperscript{1},
    Richard F.\ Lyon\textsuperscript{2}
    }

\address{\textsuperscript{1}Google DeepMind, \textsuperscript{2}Google Research, \textsuperscript{3}Google Platforms \& Devices}

\begin{document}

\maketitle 

\ninept  % Uncomment to use 9pt font, which makes more space.

\begin{abstract}
Low latency models are critical for real-time speech enhancement applications, such as hearing aids and hearables. However, the sub-millisecond latency space for resource-constrained hearables remains underexplored. We demonstrate speech enhancement using a computationally efficient minimum-phase FIR filter, enabling sample-by-sample processing to achieve mean algorithmic latency of 0.32 ms to 1.25 ms. With a single microphone, we observe a mean SI-SDRi of 4.1 dB. The approach shows generalization with a DNSMOS increase of 0.2 on unseen audio recordings. We use a lightweight LSTM-based model of 626k parameters to generate FIR taps. Using a real hardware implementation on a low-power DSP, our system can run with 376 MIPS and a mean end-to-end latency of 3.35 ms. In addition, we provide a comparison with existing low-latency spectral masking techniques. We hope this work will enable a better understanding of latency and can be used to improve the comfort and usability of hearables.
\end{abstract}
\begin{keywords}
speech enhancement, low-latency, on-device, hearables 
\end{keywords}
\section{Introduction}
\label{sec:intro}
There are many real-time applications where low-latency speech enhancement is critical. For example, hearing aids and transparency mode in hearables need low end-to-end latency to reduce comb effect filtering, increase comfort, and reduce artifacts. In such applications, end-to-end latency under 2 ms is desirable~\cite{groth2004disturbance}; thus, considering hardware overhead, the algorithmic latency needs to be 1 ms or less. Another challenge is that on hearables, computing is limited by low power, and memory is limited to around 1 MB. Machine learning approaches are state-of-the-art in real-world speech enhancement products. However, their algorithmic latency is usually around 16 to 32 ms, and their size is larger than 1 MB. Recent research has shown models with causal latencies from 1 - 8 ms~\cite{wang2021deep_lstw,schroter2022low,pandey2023simple}; however, latency, memory, and compute must be further reduced to enable widespread adoption of speech enhancement for resource-constrained hearables. This work investigates achieving the lowest latency while allowing efficient computing on embedded devices for machine-learning-based speech enhancement. 

\begin{figure}[htb]
  \centering
  % Figure was created in Adobe Illustrator, open pdf to edit.
  \centerline{\includegraphics[width=8.5cm]{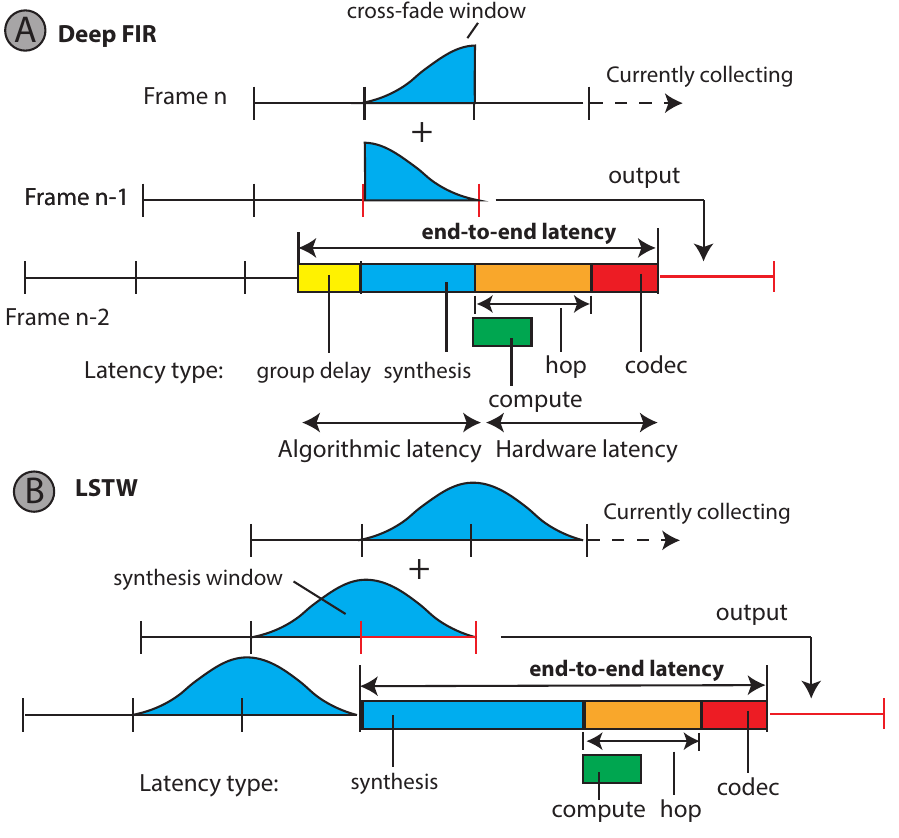}}
\caption{Visualization of end-to-end latency. A) Using the proposed Deep FIR filtering, as developed in this paper. B) Using the long-short time window (LSTW) technique. Note that the proposed Deep FIR can achieve lower end-to-end latency.}
\label{fig:diagram_latency}
\end{figure}

This paper contributes as follows: 1) A \textit{Deep FIR} method that allows algorithmic latency down to 0.34 ms with sample-by-sample processing. Audio samples can be accessed online\footnote{\href{https://google-research.github.io/sound-separation/papers/deepfir/}{google-research.github.io/sound-separation/papers/deepfir/}}. To our knowledge, this is the lowest latency demonstrated in a causal mono speech enhancement system. 2) Implementation and benchmarking on off-the-shelf low-power DSP in real-time. 3) Objective and subjective comparisons with benchmarks such as long-short time window (LSTW)~\cite{mauler2007low, wang2021deep_lstw} and common spectral mask with overlap-add (abbreviated as OLA )~\cite{wilson2018exploring} to understand latency and compute tradeoffs.

End-to-end latency has two main components, as illustrated in Figure~\ref{fig:diagram_latency}, and can be complicated due to interactions between hardware, software, and algorithms. The first part is \textit{algorithmic latency} - how much output must be delayed due to the processing algorithm, windowing, look-ahead, etc. For a causal streaming system, this delay is typically the size of the synthesis window. 
In some filters, group delay can add large algorithmic latency, such as half of the filter taps for a linear phase finite impulse response (FIR) filter. For non-linear phase filters, group delay depends on the input signal's frequencies.
The group delay in LSTW and OLA is part of the algorithmic latency.

Second is the \textit{hardware latency}, determined by the buffering, codec, and communication protocol speed between the processor and codec.
The hop size is a big part of hardware latency, which is how many samples are processed in one batch. Hop is often set as half of the analysis window. At least one hop latency is needed since hop samples can only be played once the hop is finished. The real-time model compute time is optimized to be under a hop, so processing is done while the next hop is collected. In such a case, latency would be a hop even if the compute takes less than a hop. 

The end-to-end latency can be estimated as:  \textit{synthesis window size + group delay + hop size + codec latency}. Therefore, the lowest latency system would have a 1-sample synthesis window, 1-sample hop, and minimal group delay.

 %Linear phase FIR filter can be turned into minimum phase at inference to minimize the group delay. This is done fundamentally by finding all the zeros and removing the ones outside the unit circle~\cite{herrmann1970design}.
%End to end latency = Synthesis window size + Group delay + hop + codec 
%From the latency perspective, an ideal model would have a synthesis window of 1 sample, a hop of 1 sample, and a minimal group delay. However, the model would need to run on every sample. With the improvements in computation, it might be possible to make sample-by-sample in the future, so we explore this option.

\section{Background}
% \vspace{-6pt}
\label{sec:format}
The goal of the speech enhancement task is to improve the degraded quality of speech. Mainly it is done by separation of speech from other sounds such as noise and music. 
Recently, speech enhancement has been improved with machine learning approaches. A common approach is predicting a spectral mask in the short-time Fourier transform (STFT) domain and synthesizing using overlap-add of usually 50\% overlap, such as in~\cite{wilson2018exploring}. However, most of those approaches do not have low latency and do not meet the computing and memory requirements of small devices such as hearing aids. An unmodified STFT approach will require algorithmic latency of 16 to 32 ms, a common duration of analysis and synthesis windows. For example, TinyLSTMs~\cite{Fedorov_2020} explored how speech enhancement models can be adapted to hearing aids. However, algorithmic latency is high, as 32 ms synthesis window and 16 ms hop are used. The main limitation is that the analysis and synthesis window are the same size. A technique of using an extended analysis window and short synthesis window~\cite{mauler2007low} (referred to as LSTW) can reduce latency, as the length of the synthesis window determines the algorithmic latency. LSTW has been successfully used with a deep neural network mask estimator~\cite{wang2021deep_lstw} to achieve 8 ms latency. Another approach is to predict 1 to 2 spectral masks in the future to have zero or negative latency~\cite{wang2022stft, wilson2018exploring}. Predicting the masks with LSTMs or convolutional networks reduces generalization and denoising ability.
Similarly to this work, another approach is time domain filtering. This approach has been mainly employed with low-latency traditional filter estimation~\cite{lollmann2007uniform}. Using convolutional neural network estimation of FIR-like filter has shown algorithmic latency of 8 ms~\cite{schroter2022low} or estimating an infinite impulse response (IIR) filterbank ~\cite{zheng2022low} with 4 ms latency. Currently, some of the lowest algorithmic latency of 1 ms has been shown by recurrent neural network (RNN) with convolutional front-end and direct output prediction~\cite{pandey2023simple} on multi-microphone audio.
Previous works have not explored the causal sub-1 ms latency space and their deployment on low-power systems under 1 MB memory constraints. Also, only non-causal sample-by-sample denoising has been demonstrated~\cite{luo2020dual}.

\begin{figure}[htb]
  \centering
  \centerline{\includegraphics[width=8.5cm]{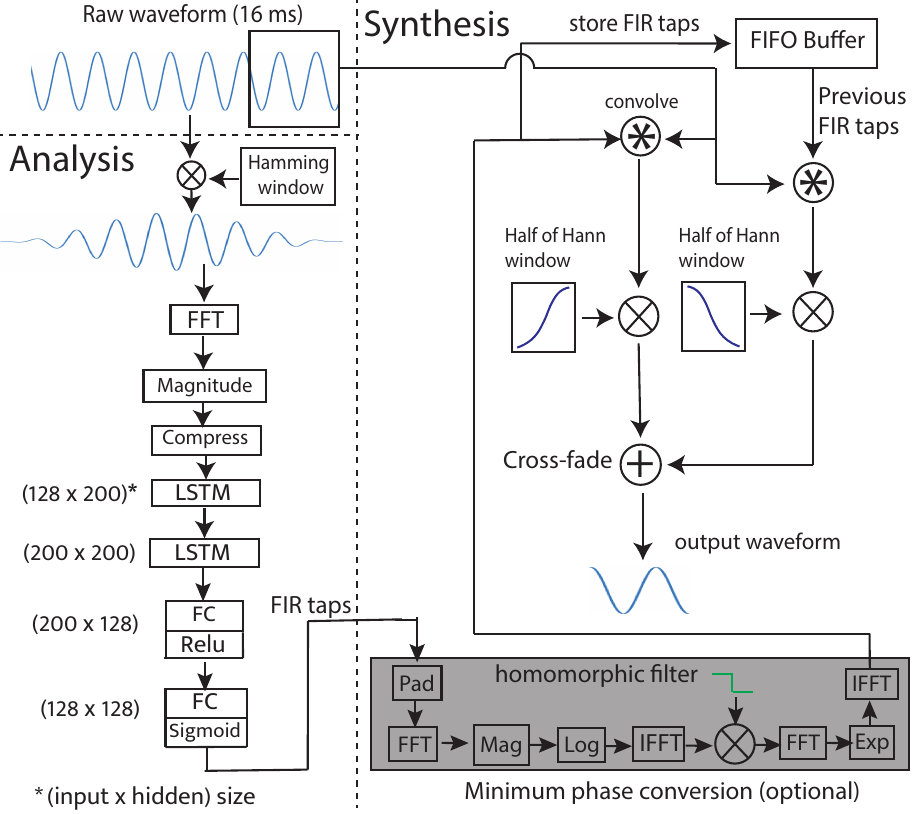}}
\caption{Inference time causal Deep FIR signal processing diagram, divided into synthesis and analysis. A new FIR filter is estimated every hop.}
\label{fig:diagram latency.}
% \vspace{-10pt}
\end{figure}

% \vspace{-12pt}
\section{Approach}
\label{sec:implementation}
\subsection{LSTM-based filter prediction network}
We use a 16 ms Hamming analysis window and compute efficient FFT magnitude features as input to the model. The magnitude is compressed to a power of 0.3. The Hamming window does not go to zeros at the edges, so some information about the latest samples is preserved. In contrast, the rectangular window conserves all the samples but introduces artifacts to the denoised speech.
We use a 2-layer long short-term memory (LSTM) network, as shown in Figure~\ref{fig:diagram latency.}. Each layer uses LSTMs with 200 units. Such an architecture has shown to be effective for speech enhancement on small embedded platforms~\cite {Fedorov_2020}. The LSTM is followed by a fully connected layer with 128 outputs and ReLU activation, followed by a fully connected layer with 128 outputs and sigmoid activation that predicts a 128-tap FIR filter. Continuous filter convolution requires 127 historic raw audio samples before the synthesis window, which are pulled from the audio segment used for analysis.

We used the compressed spectral loss function, which has been shown to work for speech enhancement applications~\cite{wisdom2019differentiable}.
Compressed spectral loss is a weighted combination of magnitude and phase-aware complex loss:
\begin{equation}
L
=\sum_{b,f,t}
(1-\beta)\left(|\hat{S}_{b,f,t}|^{\alpha} - |S_{b,f,t}|^{\alpha}\right)^2 + \beta\left(|\hat{S}_{b,f,t}^{\alpha} - S_{b,f,t}^{\alpha}|\right)^2
\end{equation}
where $\hat{S}$ is the complex reference speech STFT and $S$ is the denoised complex STFT, where $S^{\alpha} := |S|^{\alpha}e^{j\angle S}$. $b$ is batch, $f$ is frequency, and $t$ is time. $\alpha$ is the compression constant set to 0.3. $\beta$ is a constant, determining weights of magnitude and complex loss. $\beta$ was set to 0.85, which we found to improve MOS. 

%$\hat{S}^{0.3}e^{j\angle{\hat{S}}}

We did not implicitly force the linear phase into the loss function. However, the loss was calculated between the predicted audio and target delayed by half of the taps (4 ms), producing an approximate linear phase filter. 

The trained model has 626k parameters, a float32 size of 2.39 MB, and a 16x8 quantized size of 0.58 MB. Also, 2.1 KB of working memory is required for states and activations. The model was trained with TensorFlow and quantized using a TFLite converter~\cite{jacob2018quantization}. 16 kHz audio sampling rate is used. A learning rate of 1e-4 was used.
% \vspace{-8pt}
\subsection{Synthesis with Deep FIR filter}
% \vspace{-6pt}
An FIR filter with 128 taps is used, calculated by a time-domain convolution. The new filter is estimated at every hop. More taps can increase denoising but incur computational costs and add to group delay.
As a synthesis window, we use simple 50\% crossfading between two filters to avoid discontinuity artifacts between the estimates since filter coefficients can change. Crossfading is done using two halves of a Hann window, as shown in Figure~\ref{fig:diagram_latency}A. This crossfading was approximated in training with a zero-padded Hann window and 50\% overlap add after applying the FIR filter. 

Optionally, we convert the FIR filter to the minimum phase after inference to reduce the group delay. There are multiple approaches to doing so. Direct roots factorization~\cite{chen1986design} is the most accurate but is computationally demanding. In our approach, we apply a homomorphic filter to the FIR taps, which mostly uses FFTs and IFFTs as shown in Figure~\ref{fig:diagram latency.}, and is fast to compute given fast FFT ops on the DSP. This approach was adapted from ~\cite{oppenheim1999discrete}. Phase modification can affect the crossfading; however, since this approach runs with short hops, the phase does not change significantly from hop to hop. Although the model could be conditioned to predict a minimum phase filter during training, we apply it after inference to still use phase-sensitive reference-based metrics. 

% \vspace{-12pt}
\section{Results}
\label{sec:results}
% \vspace{-6pt}
\subsection{Objective evaluation of speech enhancement}
% \vspace{-4pt}
We train on the CHiME-2 WSJ0 dataset~\cite{vincent2013second}. This dataset is composed of target speech mixed with various noises recorded in a home environment. The dataset has 7138 training, 409 development, and 330 test examples at 6 SNR levels from (-3dB to +9dB). The clips had a 16 kHz sampling rate and were 3 seconds long in training.
We evaluate the CHiME-2 WSJ0 test dataset~\cite{vincent2013second} with a total of 1980 examples. Since we could not find mono baseline models in sub-2 ms latency and under 1M parameters for meaningful comparison, we trained LSTW and OLA models. For LSTW, we replicated analysis/synthesis window pair type II from~\cite{mauler2007low}. In OLA, 16 ms Hamming analysis and inverse Hamming synthesis window were used. LSTW hop was set to the half of the synthesis window and in OLA to 4 ms. All models had the same number of parameters and architecture but different algorithmic latencies.  
As metrics, we used the signal-invariant signal-to-distortion ratio (SI-SDR)~\cite{le2019sdr}, short-time objective intelligibility (STOI), perceptual speech quality objective listener (ViSQOL)~\cite{hines2015visqol}, and reference-free Deep Noise Suppression mean opinion score (DNSMOS)~\cite{reddy2021dnsmos}. 

As shown in Table~\ref{table:chime_eval} all the metrics show Deep FIR improvement over unprocessed samples for the synthesis window from 0.0625 to 1 ms. Using a longer synthesis window doesn't improve Deep FIR, as it becomes harder to estimate FIR taps.  In contrast, the LSTW approach performs worse than deep FIR for all metrics under 1 ms latency. Mask estimation is done on a large analysis window to filter a small synthesis window, which becomes harder to estimate as the synthesis window becomes smaller. We trained and evaluated models on a range of taps: 8, 16, 32, 64, 128, 228. We saw a modest improvement in noise reduction metrics as the number of taps increased. Specifically, SI-SDR increased by 0.2 from 8 to 128 taps. The effect on inference was small, affecting only the last linear layer of the model. The group delay of the linear phase FIR filter changed linearly with the number of taps (taps / 2), however, minimum phase filter conversion had a bigger impact on group delay.

% \vspace{-9pt}
\begin{figure}[htb]
  \centering
  \centerline{\includegraphics[width=8.5cm]{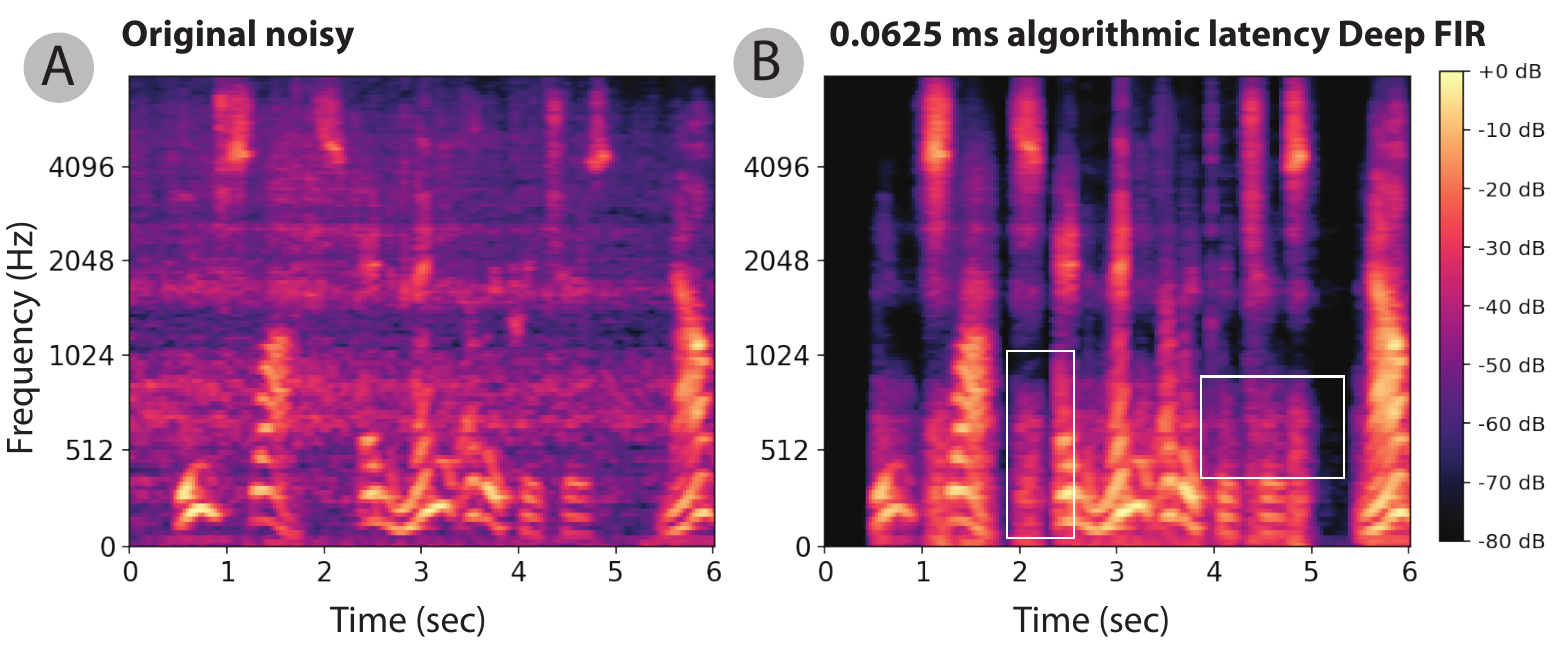}}
\caption{Example data before and after denoising. Mel spectrograms of A) original with white noise and B) speech denoised with a low-latency filter.  White rectangles point to problematic areas where the noise was not removed by the model.}
\label{fig:spectrogram_compare}
\end{figure}
% \vspace{-9pt}
The model trained on CHiME-2 provides a comparison with previous works. However, it does not generalize since it contains specific kinds of noise and speech. To investigate generalization and performance with real recordings, we trained the model on filtered speech clips from LibriSpeech~\cite{panayotov2015librispeech} and noise from Freesound~\cite{fonseca2017freesound} with SNR between -10 to +20 dB. The evaluation was done on unseen 600 clips from Deep Noise Suppression Challenge 2~\cite{reddy2021interspeech} blind test set, containing reference-free real recordings. As shown in Table~\ref{table:dnsmos_generalization}, the DNSMOS shows a consistent increase compared to unprocessed audio. In contrast, the LSTW model shows OVRL quality decrease at 0.5 ms latency. 

The primary source of noise is leakage of noise during speech. The latest samples are most important for filter estimation; however, they have the most spectral uncertainty due to window edge effects. The low frequencies are the most challenging. As shown in Figure~\ref{fig:spectrogram_compare}, when comparing the denoising with a low-latency filter and the baseline, low frequencies are also not filtered.

%\begin{figure}[htb]
%  \centering
%  \centerline{\includegraphics[width=8.5cm]{chart_mips_latency.pdf}}
%\caption{latency and computational requirements. The compute cost is benchmarked on HiFi DSP.}
%\label{fig:latency_vs_mips}
%\end{figure}

\begin{table*}[]
\centering
\footnotesize	
\captionsetup{font=small}
\caption{Model performance, trained on CHiME-2 and evaluated on CHiME-2 test set. A mean of 6 SNRs from -6 to 9 dB is reported. All models have a 16 ms analysis window.  For Deep FIR, the mean algorithmic and end-to-end latency is shown with minimum phase conversion.}
\label{table:chime_eval}
% \vspace{-5pt}
\begin{tabular}{lllllllllll}
\hline
\begin{tabular}[c]{@{}l@{}}Mean alg.\\ latency (ms)\end{tabular} &
  Type &
  \begin{tabular}[c]{@{}l@{}}Synthesis \\ win  (ms)\end{tabular} &
  \begin{tabular}[c]{@{}l@{}}Hop \\ (ms)\end{tabular} &
  \begin{tabular}[c]{@{}l@{}}SI-SDR\\ (dB)\end{tabular} &
  \begin{tabular}[c]{@{}l@{}}STOI\\ (\%)\end{tabular} &
  ViSQOL &
  \begin{tabular}[c]{@{}l@{}}DNSMOS \\ OVRL\end{tabular} &
  \begin{tabular}[c]{@{}l@{}}MIPS \\ DSP\end{tabular} &
  \begin{tabular}[c]{@{}l@{}}Can run on DSP\\ in real-time?\end{tabular} &
  \begin{tabular}[c]{@{}l@{}}Mean end-to-end \\ latency (ms)\end{tabular} \\ \hline
-    & Mix  & -  & -    & 2.31  & 72.1  & 1.35 & 1.3  & -     &     & -     \\ \hline
1    & LSTW & 1  & 0.5     & 8.0     & 76.4     & 2.23    & 2.13    & 840   & No & 2.5*   \\%\hdashline
2 & LSTW  & 2  & 1     & 11.10  & 77.40  & 2.19 & 2.25 & 420   & Yes & 4.1  \\
4    & LSTW & 4 & 2      & 11.43 & 78.49 & 2.18 & 2.28 & 210   & Yes & 7.5   \\ 
16   & OLA  & 16  & 4    & 12.98 & 80.25 & 2.35 & 2.35 & 105   & Yes & 21.2  \\ \hline
0.32 & \multirow{5}{*}{\begin{tabular}[c]{@{}l@{}}Deep FIR \\ (Proposed)\end{tabular}}   & 0.0625 & 0.0625  & 6.33     & 75.5     & 2.23    & 2.02    & 5293 & No  & 1.48* \\
0.38 &    & 0.125 & 0.125 & 6.28  & 75.6  & 2.25 & 2.01 & 2646 & No  & 1.6* \\
0.5  &    & 0.25 & 0.25  & 6.2   & 76.1  & 2.26 & 2.0  & 1359 & No  & 1.85*  \\
0.75 &    & 0.5 & 0.5   & 6.28  & 75.9  & 2.26 & 2.01 & 704  & No  & 2.35* \\%\hdashline
1.25 &    & 1  & 1    & 6.21  & 76.1  & 2.26 & 2.01 & 376   & Yes & 3.35  \\ \hline
\multicolumn{8}{l}{\footnotesize{* indicates estimated values, as those models cannot meet compute for HiFi4 DSP in real-time.}}
\end{tabular}
% \vspace{-12pt}
\end{table*}

\begin{table}[]
\centering
\footnotesize
\captionsetup{font=small}
\caption{DNSMOS on out-of-domain Deep Noise Suppression Challenge 2 dataset and trained on Freesound and LibriSpeech.}
\label{table:dnsmos_generalization}
% \vspace{-5pt}
\begin{tabular}{lllll}
Model    & \begin{tabular}[c]{@{}l@{}}Synthesis win (ms) \end{tabular} & DNSMOS : OVRL & SIG & BAK \\ \hline
mix      & -                                                      & 2.32   & 3.15 & 2.71     \\ \hline

LSTW     & 1            & \textbf{2.51} & 2.83 & \textbf{3.79}        \\
LSTW     & 0.5         & 2.29 & 2.61 &3.70           \\ 
OLA      & 16                 & \textbf{2.49}     & 2.77 & \textbf{3.85} \\ \hline
\multirow{2}{*}{\begin{tabular}[c]{@{}l@{}}Deep FIR \\ (Proposed)\end{tabular}} & 0.125       & \textbf{2.42} & 2.76 & \textbf{3.71}        \\
 & 1             & \textbf{2.52} & 2.84 & \textbf{3.78}          \\ \hline
\end{tabular}
\end{table}
%\vspace{-26pt}
\subsection{Subjective evaluation}
% \vspace{-8pt}
We conducted a subjective test according to ITU-T Rec P.835 standard~\cite{recommendation2003subjective} using 8 human subjects, all audio experts. 
A primary motivation for this test was to understand the quality of the minimum phase filter since metrics such as phase modifications make SI-SDR inaccurate. We picked a random subset of the CHiME-2 test set with 27 samples. Each sample contained either one or a mixture of interfering speech, music, and non-stationary (e.g., footsteps) or stationary (e.g., fan) environmental sounds. The set included three SNR levels: -3, +3, +9 dB SNR. Three categories were used: (1) unprocessed mix, (2) processed by 0.25 ms synthesis window minimum phase, and (3) linear phase models. T-test using SciPy~\cite{scipy_t_test} was done to calculate statistical significance ($P$ value).

Both minimum phase and linear phase processing showed a statistically significant ($P<0.01$) increase in overall quality (OVRL) and background noise (BAK) reduction compared to the unprocessed mix. However, the speech distortion (SIG) increased in the denoised samples. Minimum phase and linear phase processing did not show statistically significant differences ($P>0.01$) in any metric.  The linear phase had a higher overall quality of $3.41$ versus $3.19$ for the minimum phase. Meanwhile, the minimum phase had slightly better background noise reduction and speech distortion than the linear phase.
%\vspace{-10pt}
% MOS scores for 
% Unprocessed 
% 0.0625 ms model min phase
% 0.0625 ms model lin phase
% LSTW

% \vspace{-8pt}
\subsection{Computation and latency on hardware}
% \vspace{-7pt}
We benchmark the system on real-time performance on low-power audio DSP, which is designed for wearable devices. Specifically, we use i.MX RT600 (NXP Semiconductors) processor, which contains a 300 MHz M33 (ARM) microprocessor and 600 MHz HiFi4 Audio DSP (Cadence) in one chip. To minimize the processing time, we created a custom C inference library utilizing the HiFi NatureDSP library~\cite{nature_dsp}.

We benchmarked with a 1 ms algorithmic latency model since that is the lowest latency model we could run in real-time due to computing and hardware limitations. We achieved an end-to-end latency of 7.1 ms. Millions of instructions per second (MIPS) was measured at 304. Processing took 0.72 ms during each hop, with inference taking 0.54 ms and front-end and synthesis 0.18 ms. Hardware latency was 1.1 ms.
The minimum phase calculation added 0.043 ms to each inference and increased MIPS by 72 to 376. However, the group delay decreased from a constant 4 ms to the mean of 0.25 ms, as measured from the mean group delay at each inference of noisy speech samples. Therefore, end-to-end latency decreased to 3.35 ms.

With the same inference models and front end, the main compute difference between the Deep FIR and LSTW is in synthesis. Roughly the same algorithmic latency can be achieved with half of the hops for Deep FIR, as it does not use overlap-add. 
The Deep FIR approach requires computing a dot product between filter coefficients and raw audio for each output sample. It is efficient for a small number of samples but scales linearly with more samples.  
With the LSTW approach, compute time is constant, with dependence on the size of the analysis window. 

\begin{table}[]
\centering
\footnotesize
\captionsetup{font=small}
\caption{Subjective MOS evaluation on CHiME-2 test set comparing minimum versus linear phase for Deep FIR with 0.125 ms synthesis window.}
%\vspace{-5pt}
\begin{tabular}{llll}
MOS   & OVRL  & SIG & BAK \\ \hline
Mix              &  1.75  &  3.90   &  1.89   \\
Min phase FIR    &  \textbf{3.19 *}   &  3.18  &  \textbf{2.86 *} \\
Linear phase FIR &  \textbf{3.41 *}&  2.84  &  \textbf{2.63 *} \\
 \hline
\multicolumn{4}{l}{\footnotesize{* indicates $P<0.01$ in comparison to unprocessed mix.}}
%\vspace{-2pt}
\end{tabular}
% \vspace{-14pt}
\end{table}

% \vspace{-13pt}
\section{Conclusion}
% \vspace{-8pt}
In this paper, we demonstrated how to achieve algorithmic latency under 1 ms using an RNN model to estimate minimum phase FIR filters. We deployed such models on low-power DSP, which is suitable for hearables. We find two main limitations in deploying very low latency systems, which are opportunities for future work. First, low latency denoising will cause noise leakage during speech. Fundamentally, a certain amount of prediction is required, as due to windowing, there is some uncertainty in the frequency estimation. Second, computing the model at every hop is wasteful for short hops. Potentially, a part of the model needs to be recomputed. 

As our hearables improve, achieving minimal latency will become critical to their usability. We hope this paper provides insights into ways to achieve this.

\section{Acknowledgment} 
The authors would like to thank Leonardo Kusumo, Kevin Wilson, Pascal Getreuer, Vivek Kumar, and Jeremy Thorpe for their valuable discussions and suggestions.
% -------------------------------------------------------------------------
\bibliographystyle{IEEEbib}
%\bibliography{refs}
%{\small \bibliography{refs}} 

\small
\begin{thebibliography}{10}

\bibitem{groth2004disturbance}
Jennifer Groth and Morten Birkmose,
\newblock ``Disturbance caused by varying propagation delay in non-occluding
  hearing aid fittings,''
\newblock {\em International Journal of Audiology}, vol. 43, no. 10, pp.
  594--599, 2004.
\bibitem{wang2021deep_lstw}
Shanshan Wang, Gaurav Naithani, Archontis Politis, and Tuomas Virtanen,
\newblock ``Deep neural network based low-latency speech separation with
  asymmetric analysis-synthesis window pair,''
\newblock in {\em EUSIPCO}, 2021, pp. 301--305.
\bibitem{schroter2022low}
Hendrik Schr{\"o}ter, Tobias Rosenkranz, Alberto-N Escalante-B, and Andreas
  Maier,
\newblock ``Low latency speech enhancement for hearing aids using deep
  filtering,''
\newblock {\em IEEE/ACM Transactions on Audio, Speech, and Language
  Processing}, vol. 30, pp. 2716--2728, 2022.
\bibitem{pandey2023simple}
Ashutosh Pandey, Ke~Tan, and Buye Xu,
\newblock ``A simple rnn model for lightweight, low-compute and low-latency
  multichannel speech enhancement in the time domain,''
\newblock in {\em Interspeech}, 2023, pp. 2478--2482.
\bibitem{mauler2007low}
Dirk Mauler and Rainer Martin,
\newblock ``A low delay, variable resolution, perfect reconstruction spectral
  analysis-synthesis system for speech enhancement,''
\newblock in {\em EUSIPCO}, 2007,
  pp. 222--226.
\bibitem{wilson2018exploring}
Kevin Wilson, Michael Chinen, Jeremy Thorpe, Brian Patton, John Hershey, Rif~A
  Saurous, Jan Skoglund, and Richard~F Lyon,
\newblock ``Exploring tradeoffs in models for low-latency speech enhancement,''
\newblock in {\em IWAENC}, 2018, pp. 366--370.

\bibitem{Fedorov_2020}
Igor Fedorov, Marko Stamenovic, Carl Jensen, Li-Chia Yang, Ari Mandell, Yiming
  Gan, Matthew Mattina, and Paul~N. Whatmough,
\newblock ``Tiny{LSTM}s: Efficient neural speech enhancement for hearing aids,''
\newblock {\em Interspeech}, Oct. 2020.
\bibitem{wang2022stft}
Zhong-Qiu Wang, Gordon Wichern, Shinji Watanabe, and Jonathan Le~Roux,
\newblock ``Stft-domain neural speech enhancement with very low algorithmic
  latency,''
\newblock {\em IEEE/ACM Transactions on Audio, Speech, and Language
  Processing}, vol. 31, pp. 397--410, 2022.
\bibitem{lollmann2007uniform}
Heinrich~W L{\"o}llmann and Peter Vary,
\newblock ``Uniform and warped low delay filter-banks for speech enhancement,''
\newblock {\em Speech Communication}, vol. 49, no. 7-8, pp. 574--587, 2007.
\bibitem{zheng2022low}
Chengshi Zheng, Wenzhe Liu, Andong Li, Yuxuan Ke, and Xiaodong Li,
\newblock ``Low-latency monaural speech enhancement with deep filter-bank
  equalizer,''
\newblock {\em The Journal of the Acoustical Society of America}, vol. 151, no.
  5, pp. 3291--3304, 2022.
\bibitem{luo2020dual}
Yi~Luo, Zhuo Chen, and Takuya Yoshioka,
\newblock ``Dual-path rnn: efficient long sequence modeling for time-domain
  single-channel speech separation,''
\newblock in {\em ICASSP}, 2020, pp. 46--50.
\bibitem{wisdom2019differentiable}
Scott Wisdom, John~R Hershey, Kevin Wilson, Jeremy Thorpe, Michael Chinen,
  Brian Patton, and Rif~A Saurous,
\newblock ``Differentiable consistency constraints for improved deep speech
  enhancement,''
\newblock in {\em ICASSP}, 2019, pp. 900--904.
\bibitem{jacob2018quantization}
Benoit Jacob, Skirmantas Kligys, Bo~Chen, Menglong Zhu, Matthew Tang, Andrew
  Howard, Hartwig Adam, and Dmitry Kalenichenko,
\newblock ``Quantization and training of neural networks for efficient
  integer-arithmetic-only inference,''
\newblock in {\em CVPR}, 2018, pp. 2704--2713.
\bibitem{chen1986design}
Xiangkun Chen and Thomas~W Parks,
\newblock ``Design of optimal minimum phase fir filters by direct
  factorization,''
\newblock {\em Signal Processing}, vol. 10, no. 4, pp. 369--383, 1986.
\bibitem{oppenheim1999discrete}
Alan~V Oppenheim,
\newblock {\em Discrete-time signal processing},
\newblock Pearson Education India, 1999.
\bibitem{vincent2013second}
Emmanuel Vincent, Jon Barker, Shinji Watanabe, Jonathan Le~Roux, Francesco
  Nesta, and Marco Matassoni,
\newblock ``The second ‘chime’speech separation and recognition challenge:
  Datasets, tasks and baselines,''
\newblock in {\em ICASSP}. 2013, pp. 126--130.
\bibitem{le2019sdr}
Jonathan Le~Roux, Scott Wisdom, Hakan Erdogan, and John~R Hershey,
\newblock ``Sdr--half-baked or well done?,''
\newblock in {\em ICASSP}, 2019, pp. 626--630.
\bibitem{hines2015visqol}
Andrew Hines, Jan Skoglund, Anil~C Kokaram, and Naomi Harte,
\newblock ``Visqol: an objective speech quality model,''
\newblock {\em EURASIP Journal on Audio, Speech, and Music Processing}, vol.
  2015, pp. 1--18, 2015.
\bibitem{reddy2021dnsmos}
Chandan~KA Reddy, Vishak Gopal, and Ross Cutler,
\newblock ``{DNSMOS}: A non-intrusive perceptual objective speech quality metric
  to evaluate noise suppressors,''
\newblock in {\em ICASSP}, 2021.
\bibitem{panayotov2015librispeech}
Vassil Panayotov, Guoguo Chen, Daniel Povey, and Sanjeev Khudanpur,
\newblock ``{LibriSpeech}: an {ASR} corpus based on public domain audio books,''
\newblock in {\em ICASSP}, 2015, pp. 5206--5210.
\bibitem{fonseca2017freesound}
Eduardo Fonseca, Jordi Pons~Puig, Xavier Favory, Frederic Font~Corbera, Dmitry
  Bogdanov, Andres Ferraro, Sergio Oramas, Alastair Porter, and Xavier Serra,
\newblock ``Freesound datasets: a platform for the creation of open audio
  datasets,''
\newblock in {\em ISMIR}, 2017.
\bibitem{reddy2021interspeech}
Chandan~KA Reddy, Harishchandra Dubey, Kazuhito Koishida, Arun Nair, Vishak
  Gopal, Ross Cutler, Sebastian Braun, Hannes Gamper, Robert Aichner, and
  Sriram Srinivasan,
\newblock ``Interspeech 2021 deep noise suppression challenge,''
\newblock {\em arXiv preprint arXiv:2101.01902}, 2021.
\bibitem{recommendation2003subjective}
ITUT Recommendation,
\newblock ``Subjective test methodology for evaluating speech communication
  systems that include noise suppression algorithm,''
\newblock {\em ITU-T recommendation}, p. 835, 2003.
\bibitem{scipy_t_test}
SciPy,
\newblock ``ttest\_ind,''
  \url{https://docs.scipy.org/doc/scipy/reference/generated/scipy.stats.ttest_ind.html},
  2024,
\newblock Accessed 2024-09-03.
\bibitem{nature_dsp}
Linux Xtensa,
\newblock ``Nature{DSP} library for {HiFi4} {DSP} cores,''
  \url{https://github.com/foss-xtensa/ndsplib-hifi4}, 2023,
\newblock Accessed 2024-02-24.

\end{thebibliography}

\end{document}